\documentclass{jps-cp}
\usepackage[varg]{txfonts} 
\usepackage{color}%for modifications.

\title{Cosmological Solutions to the Lithium Problem}

\author{G. J. Mathews$^{1}$, A. Kedia$^{1}$, N. Sasankan$^{1}$, M. Kusakabe$^{2}$, Y. Luo$^{ 3,4}$, T. Kajino$^{2,3,4}$, D. Yamazaki$^{3,5}$, T. Makki$^6$, and M. El Eid$^6$}

\inst{$^{1}$Department of Physics, Center for Astrophysics, University of Notre Dame, Notre Dame, IN 46556 USA\\

$^{2}$IRCBBC$,$ School of Physics, Beihang University$,$ Beijing 100083 China\\
$^3$ National Astronomical Observatory of Japan Tokyo, 181-8588, Japan\\
$^4$Graduate School of Science, The University of Tokyo, Tokyo, 113-0033, Japan\\
$^5$University Education Center, Ibaraki University, 2-1-1, Bunkyo, Mito, 310-8512, Japan\\
$^6$Department of Physics, American University of Beirut, Lebanon}

\email{gmathews@nd.edu}

\recdate{August 26, 2019}

\abst{The abundance of primordial lithium is derived from the observed spectroscopy of metal-poor stars in the galactic halo. However, the observationally inferred abundance remains at about a factor of three below the abundance predicted by standard big bang nucleosynthesis (BBN). The resolution of this dilemma can be either astrophysical (stars destroy lithium after BBN), nuclear (reactions destroy lithium during BBN), or cosmological, i.e. new physics beyond the standard BBN is responsible for destroying lithium. Here, we overview a variety of possible cosmological solutions, and their shortcomings. On the one hand, we examine the possibility of physical processes that modify the velocity distribution of particles from the usually assumed Maxwell-Boltzmann statistics. A physical justification for this is an inhomogeneous spatial distribution of domains of primordial magnetic field strength as a means to reduce the primordial lithium abundance. Another possibility is that scattering with the mildly relativistic electrons in the background plasma alters the baryon distribution to one resembling a Fermi-Dirac distribution. We show that neither of these possibilities can adequately resolve the lithium problem. A number of alternate hybrid models are discussed including a mix of neutrino degeneracy, unified dark matter, axion cooling, and the presence of decaying and/or charged supersymmetric particles.}

\kword{Big Bang Nucleosynthesis, Cosmology, Stars and Stellar Evolution}

\begin{document}
\maketitle

\section{Introduction}

The yield of light elements from big bang nucleosynthesis (BBN) is the only direct probe of the radiation dominated epoch during the early universe. BBN occurs as the universe cools from about $10^{10}$ to $10^{8}$ K spanning times of about 1 to $10^4$ sec into the big bang. The only other probe is the spectrum of temperature fluctuations in the CMB \cite{PlanckXIII} which contains information of the first quantum fluctuations in the universe, and the details of the distribution and evolution of dark matter, baryonic matter, photons and electrons near the time of the surface of photon last scattering (about $3.8 \times 10^5$ yr into the big bang).

One of the most powerful aspects of standard BBN is the simplicity \cite{Wagoner73,Yang84,Malaney93,Iocco09,Cyburt16,Mathews17} of the equations. Because the contributions to the total mass-energy from non-relativistic matter, curvature, and dark energy are negligible, the Friedmann equation to describe the cosmic expansion is just:
\begin{equation}
\biggl(\frac{\dot a}{a}\biggr)=\sqrt{ \frac{8}{3} \pi G \rho_{\rm rad}}= 0.677 T_{\rm MeV}^2 ~ {\rm sec}^{-1} ~~,
\end{equation}
where $\rho_{\rm rad}$ is the mass energy density in relativistic particles, $H_0$ is the present value of the Hubble parameter, and $T_{\rm MeV}$ is the temperature in MeV.

Also, at the time of BBN the timescale for Compton scattering is short. Hence, the number density of particles of type $i$ with momenta between $p$ and $p + dp$ is simply given by Fermi-Dirac or Bose-Einstein distributions,
\begin{equation}
n_i (p)dp = \frac{g_i}{2 \pi^2}p^2 \biggl[ \exp\biggl(\frac{E_i(p) - \mu_i}{kT}\biggr) \pm 1 \biggr]^{-1} dp
\label{ndens}
\end{equation}
where $E_i(p)$ is the energy of the particle, the $\pm$ sign is negative for bosons and positive for fermions, while $g_i$ is the number of degenerate spin states of the particle (e.g.~$g = 1$ for neutral massless leptons, and $g_i = 2$ for charged leptons and photons).

The nuclear reactions, however, must be followed in detail as they fall out of equilibrium. For nuclide $i$ undergoing reactions of the type $i + j \leftrightarrow l + k$ one can write \cite{Wagoner73}:
 \begin{equation}
 \frac{dY_i}{dt} = \sum_{i,j,k} N_i \biggr( \frac{Y_l^{N_l} Y_k^{N_k}}{N_l ! N_k ! } n_k \langle \sigma_{l k} v \rangle - 
 \frac{Y_i^{N_i} Y_j^{N_j}}{N_i ! N_j ! } n_j \langle \sigma_{i j} v \rangle\biggr)
\label{nucrates}
 \end{equation}
where $Y_i = X_i/A_i$ is the abundance fraction, $N_i$ is the number of reacting identical particles, $n_i$ is the number density of nucleus $i$ and $\langle \sigma_{i j} v \rangle $
denoted the maxwellian averaged reaction cross section,
\begin{equation}
\langle\sigma_{i j} v\rangle=\sqrt{\frac{8}{\pi\mu_{i j}}}\left(T\right)^{-\frac{3}{2}}\int_{0}^{\infty}\sigma_{i j}\left(E\right)\: \exp{[-E/T]}\: E\: dE~~,
\label{eq:Forward Reaction Rate}
\end{equation}
where $\mu_{i j} = m_i m_j/(m_i + m_j)$ is the reduced mass of the reacting system.

Once the forward reaction rate is known, the reverse reaction rate can be deduced from an application (cf. Ref.~\cite{Mathews11}) of the principle of detailed balance. The reaction rates relevant to BBN have been conveniently tabularized in analytic form in several sources \cite{Cyburt10,CF88,NACRE}. These rates are a crucial ingredient to BBN calculations. In all there are only 16 reactions of significance during BBN. \cite{Iocco09,Cyburt16,Foley17}. Ideally, one would like to know these relevant nuclear reaction rates to very high precision ($\sim 1$\%). Fortunately, unlike in stars, the energies at which these reactions occur in the early universe are for the most part directly accessible in laboratory experiments. Although considerable progress has been made \cite{Cyburt16, Foley17,Coc17,Nakamura17,Descouvemont04,Pitrou18} toward quantifying and reducing uncertainties in the relevant rates, improved reaction rates are still desired for the neutron life time \cite{Serebrov10,Mathews05}, along with the $^2$H$(p,\gamma)^3$He, $^2$H$(d,n)^3$He, $^3$He$(d,p)^{4}$He, $^3$He$(\alpha,\gamma)^7$Be, and $^7$Be$(n,\alpha)^4$He reactions (see \cite{Ishikawa19}).

One of the powers of standard-homogeneous BBN is that once the reaction rates are known, all of the light-element abundances are determined in terms of the single parameter, the baryon-to-photon ratio, $\eta$. The crucial test of the standard BBN is, therefore, whether the independently determined value of $\eta$ from fits to the CMB reproduces all of the observed primordial abundances. Most of the best available abundance constraints have been summarized recently in \cite{Cyburt16}. Of most relevance to this work is the primordial abundance of $^7$Li.

The good news, is that once the value of $\eta$ was fixed by measurements of the CMB\cite{PlanckXIII} to be $\eta \equiv {n_b}/{n_\gamma}, \approx 2.738 \times 10^{-8} \Omega_b h^2 = (6.11 \pm 0.04) \times 10^{-10}$, there appears to be good agreement between the predictions of BBN for most light elements (i.e. D, $^3$He, $^4$He) and the primordial abundances as inferred from observations. In particular, the uncertainties in the $^4$He abundance deduced from line-strength observations of H {II} regions in low-metallicity irregular galaxies is now better understood \cite{Aver10} and agrees with BBN. Also, the D/H abundance seems very well determined from narrow-line absorption features along the line of sight to distant quasars \cite{Cooke14} and agrees surprisingly well with BBN.

\section{Lithium Abundance}
Unlike the other light elements, the primordial abundance of $^7$Li is best determined from old metal-poor halo stars with masses from about 0.75 M$_\odot$ to 0.95 M$_\odot$ and temperatures of about 6,000 K to 6,700 K corresponding to the Spite plateau (see Refs.~\cite{Iocco09,Coc17,Cyburt16} and Refs. therein). There is, however, an uncertainty in this determination due to the fact that the surface lithium in these stars may have experienced gradual depletion over the stellar lifetime due to mixing with the higher temperature stellar interiors where $^7$Li would be destroyed. The best current limit as summarized in Ref.~\cite{Cyburt16} is:
$
 ^7{\rm Li/H} = (1.58 \pm 0.35) \times 10^{-10} ~~.
 $

There is, however, one glaring problem that remains in BBN. The calculated and observed $^7$Li/H ratios differ by about a factor of 3. This is known as "the lithium problem." A number of recent papers have addressed this problem \cite{Cyburt16,Coc17,Nakamura17, Kusakabe17,Sato17,Yamazaki14,Yamazaki17,Bertulani13,Hou17,Makki19}. At present it is not yet known if this discrepancy derives from a destruction of lithium on the old stars used to deduce the primordial lithium abundance, or if it requires exotic new physics in the early universe \cite{Coc17,Nakamura17,Kusakabe17,Sato17,Yamazaki17}, or even a modification of the particle statistics in BBN itself \cite{Bertulani13,Hou17}. In this paper, we summarize some recent work and their prospects for solving the lithium problem.

%\begin{figure}[tbh]
%\includegraphics[scale=0.5]{crter.pdf}
%\caption{BBN abundances as a function of the baryon to photon ratio. Shaded bands correspond to the $2 \sigma$ (95\% C.L.) uncertainties deduced in 
%\cite{Foley17}. 
%Horizontal lines show range of the uncertainties in the primordial abundances inferred in 
%[12]. 
%Vertical lines indicate the value of $\eta$ deduced in the {\it Planck} analysis %\cite{PlanckXIII}.%\cite{PlanckXX}
%}
%\label{fig:abund}
%\end{figure}

%From Figure \ref{fig:abund} it is clear that 

\section{Solutions to the Lithium Problem?}

\subsection{Nuclear Solution}
One possible solution is in nuclear physics, such as we heard at this workshop \cite{Hayakawa19,Ishikawa19}. During the big bang, most of the lithium is formed as $^7$Be. Hence, a means to destroy lithium might be a strong resonance in the $^7$Be$(n,p)^7$Li reaction followed by the destruction of more fragile $^7$Li, and/or resonances in the $^7$Be$(n,\alpha)^4$He reaction. However, it is already clear \cite{Hayakawa19,Ishikawa19} that these resonances help but do not solve the problem.

\subsection{Astrophysical Solution}
The first author of this manuscript suspects that the most likely solution is from stellar astrophysics. It has been suggested for years, however, that the lack of a dispersion in the abundances of different stars in the Spite plateau argues against stellar destruction. This is because the star-to-star variations of stellar parameters such as rotation, meridional mixing, magnetic fields, turbulence, etc. among stars could lead to dispersions in the observed surface abundances. Nevertheless, the apparent metallicity dependence in the Li abundance \cite{Melendez10} suggests that at least some processing of lithium on the surface of these stars has occurred. Indeed, there are at least two recent works \cite{Richard05,Fu15} demonstrating that a narrow dispersion can result even after destroying lithium by a factor of 3 by turbulent diffusion \cite{Richard05} or convective over-shoot \cite{Fu15} for a broad range of stellar parameters.

\subsection{Cosmological Solutions}
Nevertheless, a number of works have looked at possible interesting cosmological solutions that involve modifications to the fundamental assumptions of the big bang itself. Here, we will discuss a few possibilities that we have considered. These illustrate the difficulty in resolving the lithium problem this way.

\subsubsection{Modified Statistics}
Although a simple Maxwell-Boltzmann (MB) distribution for the baryons is a long-standing assumption during BBN, there have been a number of recent papers in which this assumption is relaxed. For example, it is known \cite{Tsalis} that the MB distribution is not a unique solution to the Boltzmann equation. Hou et al. \cite{Hou17} considered a distribution function of the form:
\begin{equation}
 f_q({\bf v}_i) =B_q(m_i c^2/kT) \left( \frac{m_i}{2 \pi k T} \right)^{3/2}
 \left[1 -\left( q -1 \right) \frac{m_i v_i^2}{2k T} \right]^{1/(q-1)},
 \label{eq12}
\end{equation}
where $B_q(m_i c^2/kT)$ is a normalization constant determined from the requirement $\int f_q({\bf v}_i) d {\bf v}_i =1$, and $q$ is a free parameter. In the limit $q \rightarrow 1$ the MB distribution is recovered. However, by directly inserting this form into the reaction rate formula [\ref{eq:Forward Reaction Rate}] it was shown in \cite{Hou17} that the lithium abundance could be reduced enough to resolve the lithium problem. The reason this works is that for slightly positive $q$ the high-energy tail of the distribution is suppressed relative to MB statistics. Since the formation of $^7$Be via the $^3$He$(\alpha,\gamma)^7$Be reaction has the highest Coulomb barrier during BBN, it is the most sensitive to the high-energy tail of the distribution. Hence $^7$Be production is diminished.

However, this occurred at the expense of increasing the D/H value above that consistent with observations. Moreover, in subsequent work \cite{Kusakabe19} it has been shown that the assumption that the relative velocity distribution of nuclear reactions is a Tsallis distribution for individual nuclei that also obey a Tsallis distribution leads to a breakdown in momentum conservation. When this is corrected, the lithium problem cannot be resolved in this way.

\subsubsection{Primordial Magnetic Field}
One of the problems with imposing a Tsallis distribution, is that it requires a physical mechanism to force the statistics to deviate from MB. In \cite{Luo19a,Luo19b}, however, it was demonstrated that by imposing isocurvature sub-horizon fluctuations in a primordial magnetic field (PMF), the averaging over the domains after nucleosynthesis leads to an effective distribution similar to a Tsallis distribution but for which momentum conservation is implicit. However, just as in \cite{Hou17} the destruction of lithium is always at the cost of increasing the D/H abundance, and hence, a PMF is not a viable solution to the lithium problem.

\subsubsection{Relativistic Electron Scattering in the BBN Plasma}
Although the thermodynamics of a relativistic or nonrelativistic single-component gas have been known for many decades \cite{Juttner28}, the solution of the multi-component relativistic Boltzmann equation has only recently been attempted \cite{Kremer12,Kremer13} and transport coefficients have only been deduced for the case of equal or nearly identical-mass particles. Moreover, there has been recent interest in the possibility of a modification of the baryon distribution function from that of Maxwell Boltzmann (MB) statistics, either in the form of Tsallis statistics \cite{Bertulani13, Hou17}, primordial magnetic fields \cite{Luo19a}, or as a result of scattering from a background of relativistic electrons \cite{Sasankan18} which obey Fermi Dirac (FD), rather than MB statistics (see however \cite{Sasankan19}), or small relativistic corrections to Boltzmann equation derivation of the distribution function along with effects of nuclear kinetic drag. \cite{McDermott18}. 

 In the work of Ref.~\cite{McDermott18} the starting point was the FD distribution for baryons from which corrections were deduced. However, in Ref.~\cite{Sasankan18} it was noted based upon a Langevin approximation in kinetic-energy equipartition and a Monte Carlo simulation that the momentum distribution of nuclei more closely resembled the electron momentum distribution and therefore modified statistics when the electrons were relativistic. In \cite{Sasankan19}, however, a derivation has been made of the exact solution to the relativistic Boltzmann equation without an {\it a prior} assumption of what the baryon distribution should be. We showed that the problem can be approximated as an ideal two component system of baryons immersed in a bath of relativistic electrons, for which the collision term is completely dominated by elastic scattering from relativistic electrons. We showed that in the condition of relativistic pressure equilibrium (rather than kinetic-energy equipartition) the resultant baryon distribution does indeed follow MB statistics independently of the electron distribution function. This was verified by an evaluation of the relativistic Boltzmann equation and by revised numerical Monte-Carlo simulations \cite{Kedia19}. In \cite{Sasankan18} the sampling of electrons for collision with baryons was done from the distribution function $f(v)$ where $v$ is the relative velocity in the cosmological frame. However, this did not take into account the effect of the instantaneous viscosity (i.e electrons moving opposite to the direction of motion of a baryon collide more frequently with the baryon). This was corrected by sampling from $vf(v)$, where $v$ is the relative velocity in the frame of baryon. When that correction was made, the resultant distribution obeys MB statistics even for highly relativistic electrons.

 \subsubsection{Exotica}
 There have been numerous other attempts to resolve the lithium problem \cite{Mathews17}. For example, in \cite{Kusakabe17,Kusakabe14} we showed that a next-to-lightest supersymmetric $X^-$ particle (most likely the {\it selectron}) could revise the BBN reaction network. In particular, a resonant $^7$Be$_X(p,\gamma)^8$B$_X$ reaction and $^7$Be$_X \rightarrow ^7$Li $+ X^0$ decay could lead to a depletion of $^7$Li in the final BBN abundances. However, this scenario can only work for an exceedingly narrow range of lifetimes and abundances for the $X^-$ particle, without overproducing $^6$Li or deuterium.
 
 In another work \cite{Mori19, Cheoun11} it was shown that a time-dependent quark mass could lead to a depletion of lithium. In particular, it was shown \cite{Mori19} that resonance energies and widths of $^8$Be$^\ast$ states relevant to the $^7$Be$(n,p)^7$Li could be changed thereby enhancing the destruction of BBN lithium. Unfortunately, this is accompanied by an enhanced D/H abundance which precludes the possibility of solving the lithium problem.
 
 \subsubsection{Hybrid Models}
 In view of the difficulty of consistently overproducing D/H in models that attempt to reduct the $^7$Li production, there have been some attempts \cite{Yamazaki17,Makki19} to apply hybrid models involving multiple simultaneous extensions of the standard BBN model. The essence of this approach is to use one extension to deplete $^7$Li and another to restore D/H to the observed value. For example, in \cite{Yamazaki17,Yamazaki14} it was shown that the simultaneous imposition of photon cooling after BBN, plus $X$-particle decay, plus a primordial magnetic field could satisfy the D/H and $^7$Li constraints, but at the cost of overproducing $^6$Li. Alternatively in \cite{Makki19} the right combinations of varying: the neutrino temperatures; neutrino chemical potentials; number of neutrino species; plus photon cooling; and/or a form of unified dark matter, could help to alleviate (but not completely resolve) the lithium problem.
 
 \section{Conclusion}
In summary, it is the firm opinion of the first author of this manuscript that there is no obvious cosmological solution to the lithium problem. One inevitably encounters excess deuterium or violates other abundance constraints. Although, one might argue that the D/H constraint might be relaxed, the first author is convinced that the solution must be the destruction of lithium on the surfaces of metal poor halo stars. To quote the lines of King Lear, that were published in the first lines of the foundational paper of nuclear astrophysics by Burbidge, Burbidge, Fowler and Hoyle \cite{B2FH}, {\it "It is the stars, The stars above us, govern our condition"}. Although it has been fun to work on the cosmological approaches, the first author is convinced that the lithium problem is an astrophysics issue.

\section{Acknowledgments}

Work at the University of Notre Dame supported by the U.S. Department of Energy under Nuclear Theory Grant DE-FG02-95-ER40934. One of the authors (M.K.) acknowledges support from the Japan Society for the Promotion of Science.

\end{document}